\documentclass[fleqn,usenatbib,twocolumn]{mnras}

\usepackage{newtxtext,newtxmath}

\usepackage{graphicx}

\usepackage{caption}

\usepackage{float}
\usepackage{siunitx}
\usepackage{comment}
\usepackage{changepage}
\usepackage{placeins}
\usepackage{ulem}
\usepackage{caption}
\usepackage{subcaption}
\usepackage{tabularx}
\usepackage{threeparttable}

\usepackage[T1]{fontenc}

\DeclareRobustCommand{\VAN}[3]{#2}
\let\VANthebibliography\thebibliography
\def\thebibliography{\DeclareRobustCommand{\VAN}[3]{##3}\VANthebibliography}

\usepackage{graphicx}	
\usepackage{amsmath}	

\usepackage{lineno}

\title[Dust Polarisation efficiencies towards Anti-Center Galaxy]{A Comparative Study of Dust Grain Polarisation Efficiencies in the Interstellar and Intracluster Mediums towards Anti-Center Galaxy}
\author[Bijas et al.]{
N. Bijas,$^{1,2}$\thanks{E-mail: bijasthejas@gmail.com}
Chakali Eswaraiah,$^{1}$\thanks{corresponding author: eswaraiahc@labs.iisertirupati.ac.in, Ramanujan Fellow}
Panigrahy 
Sandhyarani,$^{1}$
Jessy Jose,$^{1}$
Maheswar Gopinathan$^{3}$   \\
$^{1}$Indian Institute of Science Education and Research Tirupati, Rami Reddy Nagar, Karakambadi Road, Mangalam (PO), Tirupati 517507, India.\\
$^{2}$ Jodrell Bank Centre for Astrophysics, Department of Physics and Astronomy, University of Manchester, Oxford Road, Manchester, M13 9PL, UK.\\
$^{3}$Indian Institute of Astrophysics, Sarjapur Road, Koramangala, Bangalore 560034, India.\\
}

\date{Accepted 2024 March 09. Received 2024 March 07; in original form 2023 September 28}

\pubyear{2015}

\begin{document}
\label{firstpage}
\pagerange{\pageref{firstpage}--\pageref{lastpage}}
\maketitle

\begin{abstract}
Dust polarisation observations at optical wavelengths help understand the dust grain properties and trace the plane-of-the-sky component of the magnetic field. In this study, we make use of published optical polarisation data acquired with AIMPOL along with distances ($d$) and extinction ($A_{\mathrm{V}}$) data.  We study the variation of polarisation efficiency ($P/A_{\mathrm{V}}$) as a function of $A_{\mathrm{V}}$ in the diffuse interstellar medium (ISM) and intracluster mediums (ICM) using the already published polarisation data of six clusters. Among these clusters,  NGC\,2281, NGC\,1664, and NGC\,1960 are old; while  Stock\,8, NGC\,1931, and NGC\,1893 are young. We categorize stars towards each cluster into foreground, background, and cluster members by employing two clustering algorithms GMM and DBSCAN. 
Thus, classified field stars and cluster members are used to reveal the polarisation properties of ISM and ICM dust, respectively.  We find that the dust grains located in the diffuse ISM show higher polarisation efficiencies when compared to those located in the ICM of younger clusters. 
\end{abstract}




\section{Introduction}

Dust grains in the interstellar medium (ISM) have been known to be responsible for the observed polarisation of the background starlight. When starlight passes through the helical asymmetrically shaped dust grains in the ISM  aligned with respect to the Galactic magnetic field, it undergoes differential extinction, which results in the polarisation of the transmitted light \citep{hiltner1949b}.  Observation of this polarised light can yield the fraction of polarisation ($P$), which is an important parameter that can be used to study some key dust properties \citep[][]{davis1951, LazarianHoang2007} apart from the plane-of-the-sky component of the magnetic field, i,e, polarisation angle ($PA$). Two such important properties are polarisation efficiency ($P/A_{\mathrm{V}}$) and rate of polarisation ($P/d$), which can reveal the dust alignment efficiency and the distribution of the polarizing dust grains, respectively,  along a particular line-of-sight (LOS). The study of these properties has become crucial in revealing the presence of dust populations with different properties like polarisation efficiency, rate of polarisation,  dust grain shape, size, composition, magnetic field orientation, etc., along a LOS \citep{Bijas2022}.

To understand how the dust grains with different ranges of extinction values contribute to this observed polarisation, several studies have tried to analyse the variation of $P/A_{\mathrm{V}}$ as a function of extinction ($A_{\mathrm{V}}$) by means of a power-law fit of the form $P/A_{\mathrm{V}} \propto A_{\mathrm{V}}^{-\alpha}$, where the power-law index $\alpha$ indicates the dust grain alignment efficiency at different dust layers \citep{goodman1995,chapman2011,cashmanandclemens,Alves2014,jones2014}. Typically, the values that $\alpha$ can take ranges from $-$1 to $-$0.5 \citep[][]{goodman1992,gerakines1995grain}. While an index of $\alpha$ = 0 indicates the perfect alignment of dust grains with the magnetic field in the region, $\alpha$ = $-$1 indicates the poor alignment of dust grains \citep[see,][]{whittet2008,pattle2019}. Since the alignment of the ISM dust grains with the magnetic fields is a prerequisite for using the dust polarisation to trace the plane of the sky (POS) magnetic field orientations, the polarisation efficiency of dust grains in a region is a crucial indicator to understand if the dust polarisation technique can be used to trace the magnetic field morphology in a specific region of interest.

Similarly, the polarisation rate ($P/d$) is another important parameter 
revealing the distribution of the polarising dust grains along a LOS \citep{Bijas2022}.  
It is known that the polarisation measurement of a star can trace the polarising dust grains present up to the distance of that star, assuming uniform B-field orientation all along the LOS. Observing a significant number of foreground stars, cluster members, and background stars towards a distant stellar cluster can trace the polarisation properties of the dust present in the foreground, intracluster, and background mediums, respectively. 
If foreground dust contributes a non-negligible amount of polarisation, then its removal from that of the cluster members is necessary to reveal the properties of intracluster dust. Likewise, if there exists a negligible amount of dust between a cluster and the background star, then the polarisation of the latter can reveal the polarisation properties of intracluster dust once the non-zero foreground contribution has been taken care of. Therefore, by adding distance information to the polarisation measurements, along with careful analyses, one can quantify the relative contribution of each component in comparison to the other toward a LOS.
In addition, studying how the rate of polarisation varies as a function of distance 
can help mark the distance boundary between foreground and intracluster dust.  Such analyses can further help in revealing the polarisation efficiency of foreground and intracluster dust. Previously, several studies have shown that the fraction of polarisation ($P$) increases with distance ($d$) towards several targets, but confined to the solar neighborhood \citep[][]{eswaraiah2011, lee2018, wang2017} using polarisation data from \citet{heiles2000} and distance data from \citet{hipparchus}, respectively. 

\begin{figure*}
\centering
\includegraphics[width=5.45in,height=3.5in]{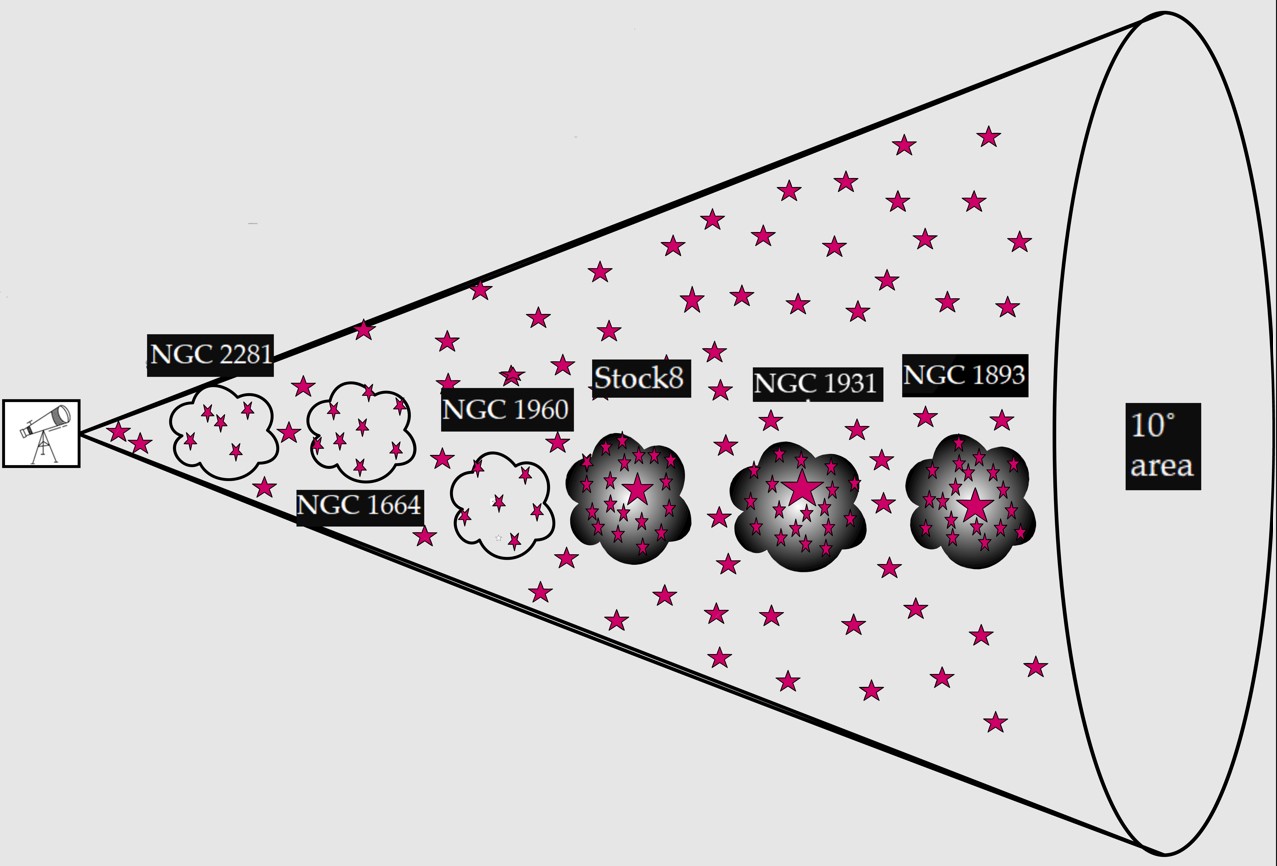}
\caption{Schematic showing the stars located towards three older clusters NGC\,2281, NGC\,1664, and  NGC\,1960 (light grey) and three younger clusters Stock\,8, NGC\,1931, and NGC\,1893 (dark) respectively. All six clusters are located at different distances but spread over a 10-degree area in the anti-center galaxy direction. The abundance of dust content is higher in younger clusters, while older clusters have a low dust content, as indicated by the dark and light grey colors of younger and older clusters. Note that the distances of clusters and the separation between them are not to scale. }
\label{three groups of stars}
\end{figure*}

In the previous work, \citet{Bijas2022} have utilized $A_{\mathrm{V}}$, distances, and polarisation measurements of stars towards NGC\,1893, and based on the analyses of $P/A_{\mathrm{V}}$ and $P/d$, they witness two dust populations. One exhibiting a higher polarisation efficiency confined to $\lesssim$ 2 kpc and another with a lower polarisation efficiency to 2 kpc.  However, in their work, they have combined field stars, cluster members, and background stars to investigate overall trends based on $P/A_{\mathrm{V}}$ ~--~ $A_{\mathrm{V}}$ and $P/d$ ~--~ $d$ plots. Combining all the stars data towards one cluster would refrain from involving dust grains of several environments with different physical conditions. In this work, we test whether there exist two dust populations even after utilizing the data of six clusters located towards the anti-center Galaxy but distributed within a large area of $10\degr$ as shown in Figure \ref{three groups of stars}. Given the fact that these clusters lie at different evolutionary stages with varying stellar activity (see Table \ref{tab:basicparameterscluster}), dust in their intracluster mediums (ICM) may exhibit different polarisation properties. In addition, since these clusters lie at different locations and distances within the Milky Way Galaxy, they may have different amounts of dust in their foregrounds and backgrounds, resulting in complex dust properties. Since all these clusters are located towards the anti-center Galaxy, by classifying the stars of each cluster into three groups such as foreground, background, and cluster members, and carefully testing the dust component that each group traces, we analyse $P/A_\mathrm{{V}}$~--~$A_{\mathrm{V}}$ trends of the ISM and ICM dust.  It is important to note that the polarisation observations for the six clusters within the 10-degree area are obtained from slightly different lines of sight (LOSs), which cover a relatively small portion of the sky. Therefore, throughout this work, we assume that the light from all the clusters passes through a similar dust content of ISM and that the properties of the dust do not significantly vary from one LOS to another.

The structure of this paper is as follows. Section \ref{section2} describes various data sets, utilised in this study,  such as optical polarisation, distance, and extinction for the six clusters. Section \ref{section3} presents the analyses for examining the presence of two dust populations towards these clusters. This is followed by the classification of stars towards each cluster into foreground, background, and cluster members using unsupervised machine learning algorithms GMM and DBSCAN. This classification helps delineate the observed polarisation contributions from ISM and ICM dust. We further detail the results based on $P/A_\mathrm{{V}}$~--~$A_{\mathrm{V}}$  relations for the ISM and ICM on the basis of power-law fits. The results are discussed in Section \ref{section4}. Finally, Section \ref{section5} summarizes our results and presents our conclusions.


\begin{table*}
\centering
\caption{The basic parameters of all the six clusters distributed in the anti-center Galaxy direction.  Note that the distances quoted in this table denote the overall distance to the stellar clusters identified from different publications.}
\label{tab:basicparameterscluster}
\begin{threeparttable}
\begin{tabular}{lccccccl}
 \hline
 Cluster-ID & Galactic longitude (l) & Galactic latitude (b) & Distance & log (age) \\
\hline
& (J2000; degree)&(J2000; degree)&(kpc)& (Myr)  \\
\hline
NGC\,2281  & 174.90  & $+$16.88  & 0.56  & 8.70  \\
NGC\,1664  & 161.68  & $-$0.45   & 1.20  & 8.72 \\
NGC\,1960  & 174.54  & $+$1.07   & 1.33  & 7.40\\
Stock\,8   & 173.37  & $-$0.18   & 2.05  & 6-6.70 \\
NGC\,1931  & 173.90   & $+$0.28  & 2.30  & 6.0 \\
NGC\,1893  & 173.59  & $-$1.68   & 3.25 & 6.60 \\
\hline
\end{tabular}
\begin{tablenotes}
\item[1] The quoted information for NGC\,1931 has been taken from \citet{pandeyetal2013} and for the rest of the clusters is from \citet{eswaraiah2011}.
\end{tablenotes}
\end{threeparttable}
\end{table*}

\section{Data}\label{section2}
In this work, we use the available data on polarisation, extinction, and distance.  The $V$-band polarisation data of 14 stars towards NGC\,2281, 27 stars towards NGC\,1664, 15 stars towards NGC\,1960, 21 stars towards Stock\,8, and  52 stars towards NGC\,1931. But for NGC\,1893, the $I$-band polarimetric observations of 152 stars are considered \citep{Bijas2022} in this analysis. 
All the data were acquired with the ARIES Imaging Polarimeter (AIMPOL) \citep[][]{rautela} mounted on the Cassegrain focus of the 104-cm Sampurnanand telescope at ARIES, Nainital, India. The details on the polarimetric observations towards NGC\,2281, NGC\,1664, NGC\,1960, and Stock\,8 can be found in \citet{eswaraiah2011}, for NGC\,1931 in \citet{pandeyetal2013}, and for NGC\,1893 in  \citet{Bijas2022}.  All the clusters, except NGC\,1893, have V-band polarisation data. To make the polarisation data from all clusters uniform, the $I$-band polarisation data of NGC\,1893 \citep{Bijas2022} were converted to $V$-band using the Serkowski law \citep{serkowski1975}
\begin{eqnarray}
P_{I} = P_{max}\exp\left[-K \ln^2\left(\frac{\lambda_{max}}{\lambda_{I}}\right)\right],
\end{eqnarray}
where $P_{I}$ is the percentage polarisation in the $I$ band ($\lambda_{I}$ = 0.88 $\micron$),  $P_{max}$ is the peak polarisation, and is assumed to occur in the V-band ($\lambda_{max}$ = 0.55$\pm$0.01 $\micron$; \citealt{eswaraiah2011}). The Serkowski parameter $K$ is estimated using the relation
$K = c_{1}\lambda_{max}+ c_{2}$, 
where $c_{1}$ = 1.66 $\pm$ 0.09 and  $c_{2}$ = 0.01 $\pm $ 0.05 are the constants for the visible to near-infrared (0.35 $\micron$ < $\lambda$ < 2.2 $\micron$) regime \citep{whittet1992book}. The uncertainties in $P_{I}$ are estimated by propagating the uncertainties in $P_{max}$, $\lambda_{max}$, and $K$. Here after, $P_{I}$ is treated as $P$.

The corresponding stellar distances are obtained from \citet{bailerjones2021}. Bailer-Jones catalogue has geometric and photogeometric distances based on the stellar parallaxes of 1.47 billion stars published in {\it Gaia} Early Data Release 3 ({\it Gaia} EDR3). Since simulated data and external validations show that photogeometric distances are a  better estimate for distant and faint stars \citep[][]{bailerjones2021}, we use photogeometric distances for our study.  The details of how the geometric and photogeometric distances are estimated can be found in \citet{bailerjones2021}.  We cross-match the coordinates of stars with polarisation data towards all the six cluster regions with the Bailer-Jones catalogue within 2.7\arcsec, 4.6\arcsec,  1.5\arcsec, 2.2\arcsec, 0.7\arcsec, and 0.5\arcsec respectively matching radii to find the photogeometric distances of all the stars towards NGC\,2281, NGC\,1664, NGC\,1960, Stock\,8, NGC\,1931, and of 151 stars towards NGC\,1893.

The total extinction in $V$-band, $A_{\mathrm{V}}$, are extracted from the three-dimensional dust reddening map published in \citet{green2019} using {\sc dustmaps} module in {\sc python}. They have used the stellar photometry from optical Pan-STARRS 1 and near-infrared 2MASS along with the {\it Gaia } EDR2 distances to infer the dust reddening values. The details of the dust reddening map and the methods and equations used for extracting $A_{\mathrm{V}}$ values used in this study are described in detail in \citet{Bijas2022}. Out of the obtained reddening values for all the stars, some stars without reliable distance information in the 3D dust reddening map have been removed. This was based on whether the LOS fit of cumulative reddening vs. distance has converged in a given LOS and also whether the distance at which the extinction values are returned is accurate. Hence, we obtain $A_{\mathrm{V}}$ values for 13 stars towards NGC\,2281, 25 stars towards NGC\,1664,  14 stars towards NGC\,1960, 21 stars towards Stock\,8, 51 stars towards NGC\,1931,  and 143 stars towards NGC\,1893, respectively. 

\section{Analyses}\label{section3}
\subsection{\textbf{Overall \textbf{$\mathbf{P}/\mathbf{A}_{\mathrm{\mathbf{V}}}$ \textbf{\textit{versus}} $\mathbf{A}_{\mathrm{\mathbf{V}}}$}  relation towards the six clusters}}\label{3.1}

In this section, we analyse how $P/A_{\mathrm{V}}$ varies as a function of $A_{\mathrm{V}}$ and $P/d$ varies as a function of $d$ towards a larger sky area of 10$\degr$ diameter consisting of all the six clusters in the anti-center galaxy direction. This would help in understanding the overall variations in polarisation properties of dust that exist in the foreground, background, and intracluster mediums.

\begin{figure}
    \captionsetup[subfigure]{labelformat=empty}
    \begin{subfigure}{\linewidth}
        \centering
        \includegraphics[width=4in,height=3.125in]{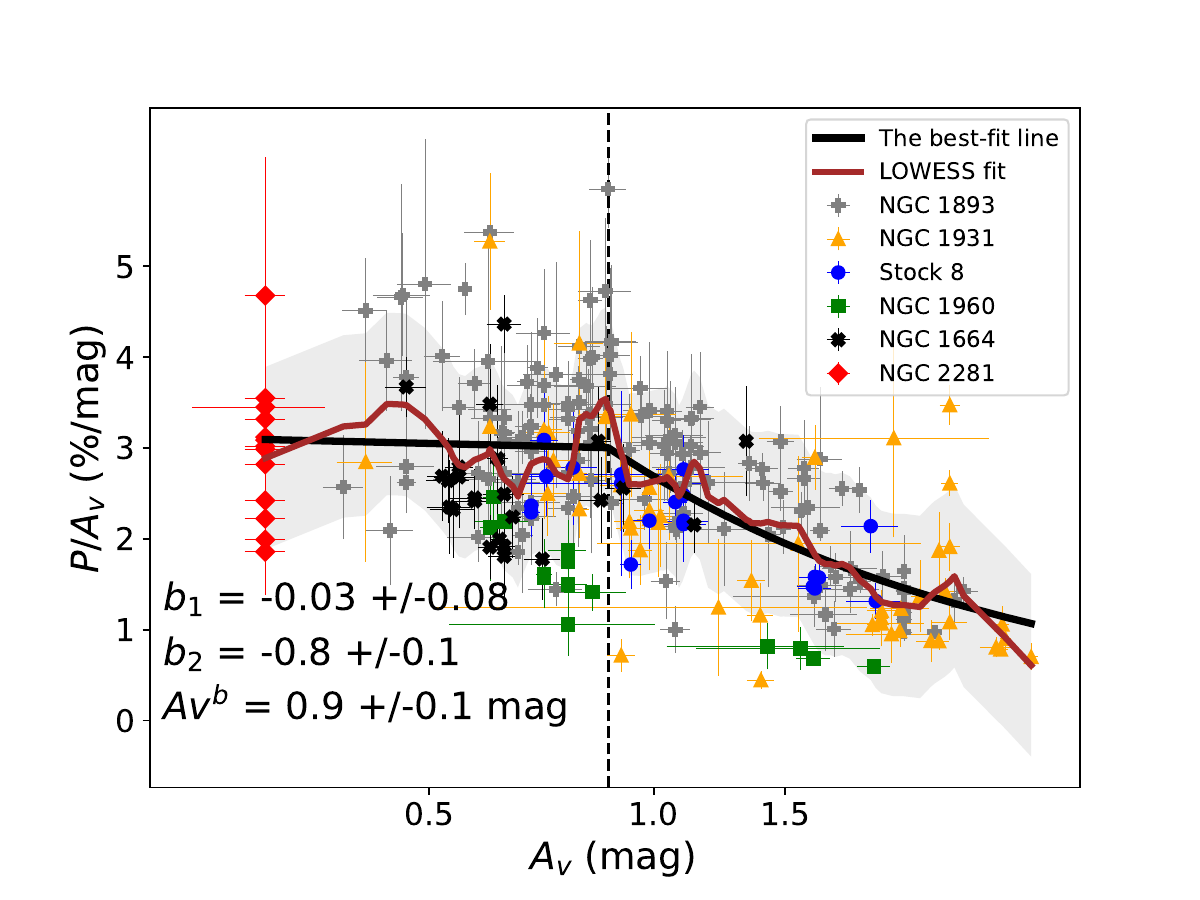}
        \caption{\label{fig:pbyavbroken}}
    \end{subfigure}\hspace{1em}
    \begin{subfigure}{\linewidth}
        \centering
        \includegraphics[width=4in,height=3.125in]{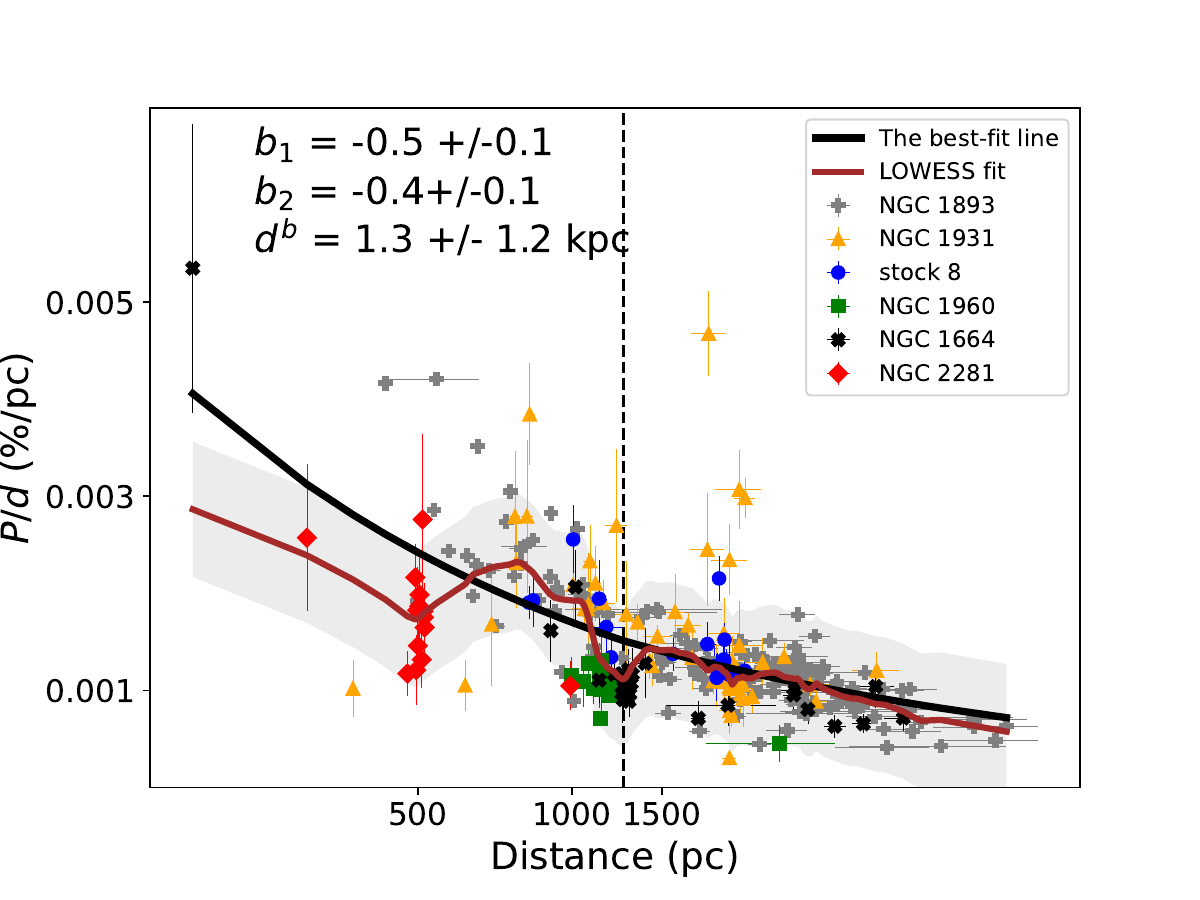}
        \caption{\label{fig:pbydbroken}}
    \end{subfigure}
    \caption{The overall $P/A_{\mathrm{V}}$ {\it versus} $A_{\mathrm{V}}$ plot ({\it top}) and $P/d$ {\it versus} $d$ plot ({\it bottom}). The best fit broken power law is denoted with a thick line. The dotted vertical line corresponds to $A_{\mathrm{V}}^b$ and $d^{b}$ at which the power law breaks and the best-fit parameters are overlaid. The overall trends traced by the LOWESS smoothed curves along with the corresponding 1-$\sigma$ regions shown as shaded areas in the background, are also displayed in both panels.}
    \label{brokenpowerlawplots}
\end{figure}

The  relation $P/A_{\mathrm{V}}$~--~$A_{\mathrm{V}}$ towards the  six clusters are fitted with a broken power law of the form using the $scipy-curvefit$ module in {\sc Python}:
\begin{eqnarray}\label{eq:brokenpowerlawav}
    y=\begin{cases}
 a\left(\frac{x}{A_{\mathrm{V}}^b}\right)^{b_{1}}, & \text{for}~~ x \leq {A_{\mathrm{V}}}^{b}\\[10pt]
a\left(\frac{x}{A_{\mathrm{V}}^b}\right)^{b_{2}}, &
 \text{for}~~ x > {A_{\mathrm{V}}}^{b},
 \end{cases}
\end{eqnarray}
where $x$ is $A_{\mathrm{V}}$,  $A_{\mathrm{V}}^b$ is the $A_{\mathrm{V}}$ at which the power-law breaks, and $a$ is a coefficient. The $b_{1}$ and $b_{2}$ are power-law indices before and after $A_{\mathrm{V}}^b$.  Similarly, the data of $P/d$~--~$d$ are also fitted with the broken power-law of the form similar to equation \ref{eq:brokenpowerlawav}, where $A_{\mathrm{V}}$ and $A_{\mathrm{V}}^b$ are replaced with distance ($d$) and distance break $d^b$ at which the power-law breaks, respectively.  Figure \ref{brokenpowerlawplots}   shows the data and corresponding best fits for $P/A_{\mathrm{V}}$~--~$A_{\mathrm{V}}$  and $P/d$~--~$d$ , respectively. The best-fit parameters are given in Table \ref{tab:brokenpowerlawparams}.

To determine whether the broken power-law model of $P/A_{\mathrm{V}}$~--~$A_{\mathrm{V}}$ and  $P/d$~--~$d$ is a better fit than a single power-law, we perform a reduced $\chi^2$ test. For $P/A_{\mathrm{V}}$~--~$A_{\mathrm{V}}$, the $\chi^2$ values for the broken power-law and single power-law models were 64.74 and 74.53, respectively. This indicates that the broken power-law model is better for $P/A_{\mathrm{V}}$~--~$A_{\mathrm{V}}$  due to its lower $\chi^2$ value. For $P/d$~--~$d$, the $\chi^2$ values for both the broken and single power-law models were 0.049, suggesting that both models fit equally well. However $P/d$~--~$d$ relation is very complex (as indicated by the LOWESS curve discussed later in this section) and has a significant amount of scatter caused due to the data points of NGC\,1931 (yellow) located beyond 1.5 kpc. Therefore, we opted for the broken power-law model as it is more capable of capturing complex trends in the data and is sensitive to the outliers when compared to single power-law.

Based on the broken power-law fitting on  $P/A_{\mathrm{V}}$~--~$A_{\mathrm{V}}$, shown in Figure \ref{brokenpowerlawplots}  ({\it Top}), we find that the polarisation efficiency ($P$/$A_{\mathrm{V}}$) remains constant up to a threshold extinction, $A_{\mathrm{V}}^b$, of  0.9 $\pm$ 0.1 mag with a power-law index of $-$0.03 $\pm$ 0.08. For $A_{\mathrm{V}} > 0.9$ mag the dust polarisation efficiency decreases with a power-law index of $-$0.8 $\pm$ 0.1. Similarly, the broken power-law over the data of $P/d$~--~$d$, shown in Figure \ref{brokenpowerlawplots} ({\it Bottom}), reveals a steeply decreasing trend by the rate of polarisation up to a break distance, $d^{b}$, of 1.3 $\pm$ 1.2 kpc by following an index of $-$0.5 $\pm$ 0.1. Thereafter, the rate of polarisation still decreases, but with a shallower power-law index of $-$0.4 $\pm$ 0.1.  In both $P/A_{\mathrm{V}}$~--~$A_{\mathrm{V}}$ and $P/d$~--~$d$ plots, the NGC\,1931 data points (yellow) lying beyond $A_{\mathrm{V}}$ $>$ 1 mag and distance $>$ 1.5 kpc is seen to exhibit significant scatter. We overlaid the location of these data points on the R-band image of the NGC\,1931 cluster and found that these stars were located near the nebulous region of NGC\,1931 \citep{pandeyetal2013}, where the O/B type stars are located. Differential reddening within the cluster and the polarization contribution from the nebulous medium could be the reasons behind the observed scatter.
Similarly, the NGC 2281 data points (red) are seen to exhibit scatter with $P/A_{\mathrm{V}}$ $>$ 1.5 \% mag in the $P/A_{\mathrm{V}}$~--~$A_{\mathrm{V}}$ plot (Figure \ref{brokenpowerlawplots} ({\it Top})). We assessed their impact on the fit by excluding them. Despite their removal, the consistency of the fit was maintained, indicating that these points have only a minimal impact on the fitting process.

From the overall trends in  $P/A_{\mathrm{V}}$~--~$A_{\mathrm{V}}$ and 
$P/d$~--~$d$ relations, we see that they match with the results from previous work, \citep{Bijas2022} in which they observed similar trends in $P$/$A_{\mathrm{V}}$ and $P/d$ towards the NGC\,1893. This confirms the existence of two dust populations: foreground and Perseus dust, the former with higher polarisation efficiency and a lower rate of polarisation and the latter with lower polarisation efficiency and a slightly higher rate of polarisation, present in a 10-degree diameter in the anti-center galaxy direction as shown in Figure \ref{three groups of stars}.

Even though the results based on the power-law fits (Figure \ref{brokenpowerlawplots}) look similar to those seen towards NGC\,1893 \citep{Bijas2022}, there is a difference in the overall data distribution in the plots of $P/A_{\mathrm{V}}$ versus $A_{\mathrm{V}}$. In the case of NGC\,1893, the data distribution is notably smooth as shown in the top panel of Figure 4 of \citet{Bijas2022}. However, when all clusters are taken into account, there are several noticeable deviations from this smooth trend, as clear from LOWESS smoothed curves (brown) shown in Figure \ref{brokenpowerlawplots}.  The curves are drawn using locally weighted scattered smoothing (LOWESS) algorithm \citep{cleveland1979} with a frac parameter of 0.1, using {\it statsmodels} package in {\sc python}. 
These variations are attributed to a well-mixed foreground/background and cluster member stars, which could trace dust polarisation properties of interstellar and intracluster mediums, respectively. Moreover, the ISM dust properties towards one LOS may differ from other LOSs. Similarly, the dust properties in the ICM of one cluster might not be similar to that of another due to the difference in the amount of the intracluster medium, number of O/B type stars, etc. Therefore, it will be highly difficult to interpret the fluctuating trends in the $P/A_{\mathrm{V}}$ versus $A_{\mathrm{V}}$ plots of various clusters. This is especially pronounced if clusters chosen have different ages, varying amounts of foreground/background, and intracluster mediums.

To facilitate a comprehensive understanding of the dust polarisation efficiency of the dust in the ISM and ICM, one has to separate the total observed stars into foreground and background stars, and cluster members. Then, by using $P$ and $A_{\mathrm{V}}$ of them, we will study the polarisation efficiencies of dust in the ISM and ICM.

\subsection{Identification of stars tracing ISM and ICM dust}\label{3.1}

To separate the observed polarisation contributions 
towards six clusters into the ISM and ICM components, it is essential to classify all the stars with polarisation and distance data into foreground, cluster, and background stars. Since the member stars of a cluster are believed to be formed by the same parental cloud, they are expected to be located at the same distances. In addition, their light will pass through similar amounts of foreground column density resulting in similar amounts of fraction of polarisation (based on the average relation between polarisation and extinction ($P = 5\,E(B-V)$ or $P = 1.6\,A_{\mathrm{V}}$; assuming $R_{\mathrm{V}} = 3.1$; \citealt{serkowski1975}). Therefore, in the polarisation versus distance plot, cluster members are expected to exhibit a conspicuous grouping or clustering.  In contrast, both foreground (which lie between the cluster and the observer) and background (lie behind the stellar cluster) stars show a scattered distribution depending upon their distances and the amount of extinction they encounter.

In addition, considering the proper motion of stars provides another crucial parameter for classifying stars into field and cluster members. This is due to the fact that all stars within a cluster being part of a gravitationally bound system can exhibit similar, albeit smaller, proper motions in comparison to the foreground field stars resulting in a clustered distribution in the proper motion plot ($\mu_{\mathrm{RA}}$ vs. $\mu_{\mathrm{Dec}}$; example see top panel of Figure 8 of \citealt{Bijas2022}).  In contrast, foreground stars, being closer than the cluster members and not part of a gravitationally bound system, will exhibit higher proper motions. Background stars, located farther than the cluster members, will display proper motions similar to those of the cluster members (see Figure 9 of \citealt{Bijas2022}). Therefore, combining the distance information of the stars with the polarisation data and proper motions offers a three-dimensional perspective on the location of each star towards a particular LOS, yielding reliable membership information. 


To separate the observed stars into cluster and field stars, we employ two unsupervised {\it {clustering}} algorithms such as the Gaussian Mixture model (GMM) for  NGC\,2281, NGC\,1960, Stock\,8, NGC\,1931, and NGC\,1893 and the Density-Based Spatial Clustering of Applications with Noise (DBSCAN) for only one cluster NGC\,1664. 
The choice of the above-mentioned unsupervised algorithms for classifying purposes depends mostly on the parameters such as  the number of mixture components ($n_{components}$) for GMM,  epsilon  ($\epsilon$) and minimum number of points ($MinPts$) for DBSCAN as well as the probability density structure of each dataset.
While in the case of the five clusters, we found the GMM-predicted clusters could trace the background density structure, in the case of NGC\,1664 it failed to do so. 
Therefore, for NGC\,1664 we decided to go with DBSCAN, the next-best algorithm that can predict clusters based on their density.

\begin{figure*}
\centering
\captionsetup[subfloat]{labelformat=empty}
\begin{minipage}{1\textwidth}
\centering
\hspace*{\stretch{1}}
\subfloat[(a)]{\includegraphics[width=0.45\textwidth]{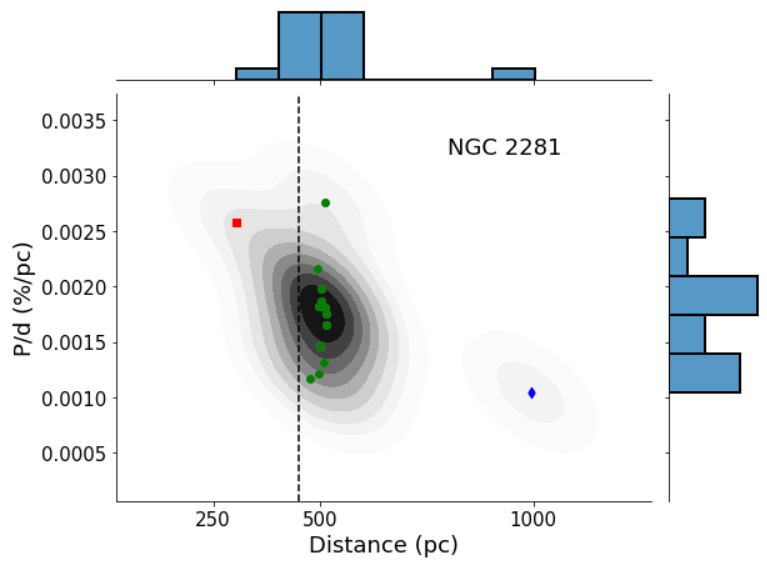}}%
\subfloat[(b)]{\includegraphics[width=0.45\textwidth]{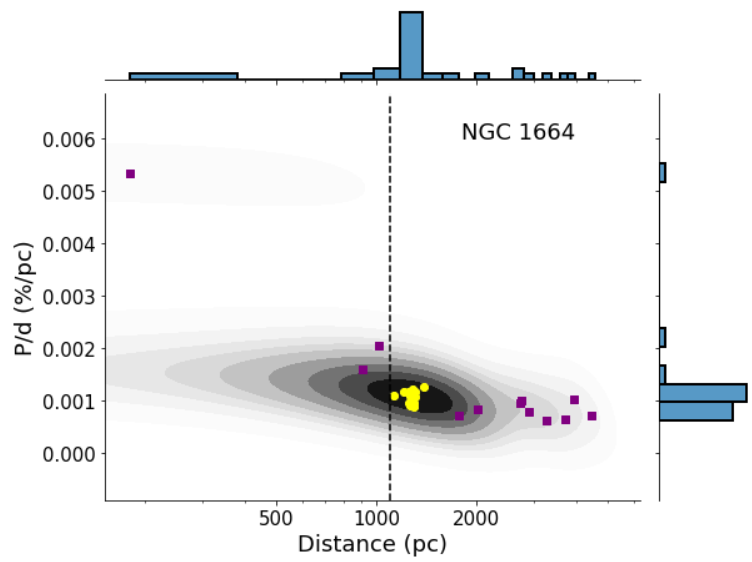}}%
\vspace{0.1cm}
\subfloat[(c)]{\includegraphics[width=0.45\textwidth]{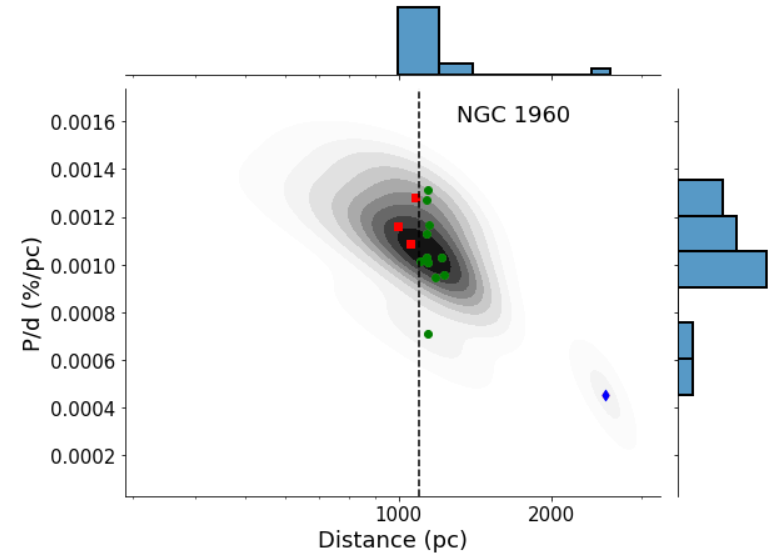}}%
\subfloat[(d)]{\includegraphics[width=0.45\textwidth]{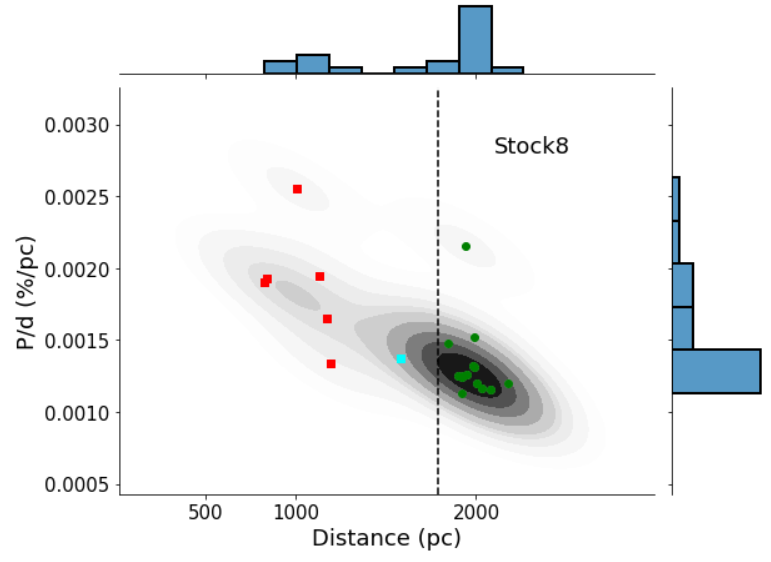}}%
\vspace{0.1cm}
\subfloat[(e)]{\includegraphics[width=0.45\textwidth]{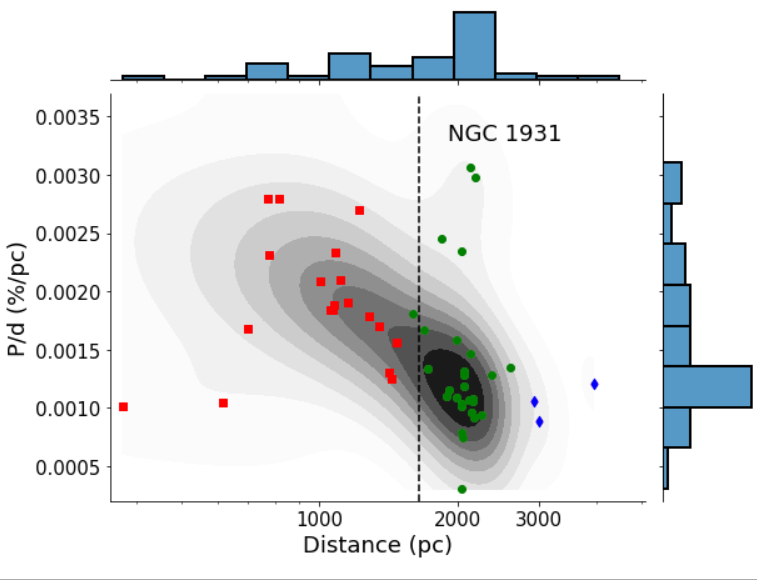}}%
\subfloat[(f)]{\includegraphics[width=0.45\textwidth]{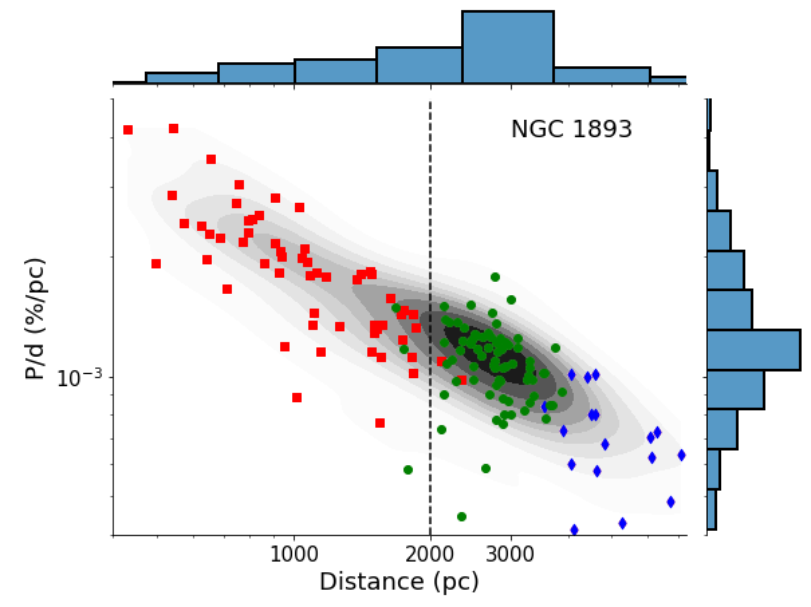}}%
\hspace{\stretch{1}}

\caption{The $P$/$d$ {\it versus} distance ($d$) relations towards each cluster. In panels  (a), (c), (d), (e), and (f), three groups of stars—foreground stars (squares), cluster members (circles), and background stars (diamonds), identified by the GMM, are overlaid. Note that  Stock\,8 (panel d) does not have background stars and has two groups of foreground stars identified by the GMM instead. Panel (b) shows the two groups of stars,  foreground stars (squares) and cluster members (circles), identified by DBSCAN. The darker background in all panels corresponds to the higher probability density, whereas the lighter density for the lower probability density. The vertical dotted lines in each figure denote the transition distance,  where the transition from ISM to ICM is observed.  One-dimensional histograms in each panel are also shown.}
\label{pbydcluster}
\end{minipage}
\end{figure*}

The GMM analyses are applied on the $P/d$~--~$d$ plots, by providing,  distance $d$, Stokes parameters $Q$ and $U$,  proper motions $\mu_{\mathrm{RA}}$ and $\mu_{\mathrm{Dec}}$ (obtained from {\it Gaia} EDR3 catalogue) and their corresponding errors as input using the $scikit-learn$ module in {\sc python}. A sample data table consisting of the input parameters, $P/d$, and $P/A_{\mathrm{V}}$ are given in Table \ref{tab:NGC1893}.  The GMM is based on the assumption that data instances come from multiple Gaussian distributions with unknown mean and covariance values \citep{melchior2018,cantat2019}. It is a useful algorithm for grouping stars according to their properties. GMM assigns a probability for each star to be part of a particular group using an Expectation-Maximization (EM) algorithm. The EM algorithm starts off by assigning group parameters to the stars and then iteratively performs two steps -- expectation and maximization, until it converges. In the expectation step, the EM algorithm calculates the probability for each star belonging to a particular group based on the current group parameters. In the maximization step, the EM algorithm updates each group based on all the stars in the group, with each star weighed by the probability that it belongs to that group. We give two parameters as input to GMM for performing the classification: (i) $n_{components}$ = 3 as we expect three groups of stars along a LOS and (ii) random state = 0 or 1 for maintaining consistency in the classification. The GMM has categorized the total sample of stars towards each stellar cluster into the foreground, cluster members, and background stars, 
which are shown in Figure \ref{pbydcluster} (a), (c), (d), (e), and (f). 
The information on probability density and frequency distributions are also shown in the figures with black/grey background and histograms, respectively.

Similarly, DBSCAN classifies cluster members and field stars towards NGC\,1664 using the $scikit-learn$ package. DBSCAN 
identifies clusters by grouping data points together in dense regions and identifies data points in low-density regions as noise or outliers \citep{ester1996,schubert2017}. It uses two input parameters, epsilon ($\epsilon$) and the minimum number of points ($MinPts$), to determine the density of the clusters. To start clustering, the algorithm randomly selects a point and searches for neighbouring points within a radius of $\epsilon$. If the number of neighbouring points is greater than or equal to $MinPts$, a cluster is formed, and the algorithm expands the cluster by adding all the points within the $\epsilon$ distance of the initial cluster. The algorithm then identifies the next unassigned point and repeats the process until all points are assigned to a cluster. It is important to note that the parameters for DBSCAN are specific to the distribution of individual datasets. In the case of the NGC\,1664 cluster,  $\epsilon$, and $Minpts$ parameters were fixed at 90 and 5, respectively. The DBSCAN-identified field and cluster members towards NGC\,1664, along with 
the probability density and frequency distributions are shown in Figure \ref{pbydcluster} (b).

The distance at which the transition from the foreground medium to the intracluster medium occurs can be determined by the fact that the number of stars and, hence, the background density increase abruptly, and thereafter, a cluster of data points appears as shown in Figures \ref{pbydcluster} (a)~--~(f). 
The clusters NGC\,1893, NGC\,1931, Stock\,8, NGC\,1960, NGC\,1664, and NGC\,2281 exhibit transition distances  (note that the {\it distance break} based on the power-law break shown in Figure \ref{brokenpowerlawplots} is different from the transition distance) of 2, 1.65, 1.79, 1.1, 1.2, and 0.45 kpc, respectively, as denoted with vertical dotted lines in Figures \ref{pbydcluster} (a)~--~(f). 

Three of the six clusters, Stock\,8, NGC\,1931, and  NGC\,1893, we analyse are young, and the other three, NGC\,2281,  NGC\,1664, and NGC\,1960 are old (see Table \ref{tab:basicparameterscluster}).  Since younger clusters contain leftover cloud material, the background stars may show polarisation properties similar to those of cluster members. So we consider stars lying beyond the transition distance to be cluster members and background stars representing ICM dust, and stars up to the transition distance to be foreground stars revealing ISM dust.  However, the old clusters are supposed to have a negligible dust content; hence, the excess amount of polarisation from the old cluster member's ICM should be near zero. As a result, the cluster members, background stars, and foreground stars along the LOSs of old clusters can trace the ISM very well.  Consequently, it is reasonable to assume we can only trace the polarisation properties of ICM dust in younger clusters and not in older clusters.

We have adopted the following assumptions for characterizing the polarisation properties of ISM and ICM dust: (a) ISM dust is traced by foreground stars of all young and old clusters, plus cluster members and background stars of old clusters (hereafter ISM tracers); (b) ICM dust is traced by both cluster members and background stars of young clusters (hereafter ICM tracers).
%
By extracting $P$, $A_{\mathrm{V}}$, and distance values of all the ISM and ICM tracers, we analyse $P/A_{V}$~--~$A_{V}$ relations separately for the dust in ISM and ICM.

\begin{table*}
\centering
\caption{Sample data table containing the Source ID, cluster name, type of tracer (ISM or ICM), equatorial coordinates,  distance (from Gaia EDR3), Stokes parameters, proper motions, polarisation rate, extinction, polarisation efficiency, 
Intrinsic extinction and intrinsic polarization efficiency, along with their errors.
}
\begin{tabular}{c c c c c c c c c }
\hline
Source\_ID & Cluster & Tracer & RA (J2000) & Dec (J2000) & Distance $\pm$ $\sigma$ & Q $\pm$ $\sigma$ & U $\pm$ $\sigma$  \\
\hline
 &  &  &  (degree) & (degree) & (kpc)  & (\%) & (\%)\\
    \hline
181173402856344704 & NGC 1893 & ICM & 80.6653 & 33.3717 & 3.00 $\pm$ 0.16 & 1.29 $\pm$ 0.20 & -2.04 $\pm$ 0.21\\
181173677734249600 & NGC 1893 & ICM & 80.7238 & 33.3921 & 2.22 $\pm$ 0.10 & 2.19 $\pm$ 0.55 & -1.09 $\pm$ 0.55 \\
 181173677734250240 & NGC 1893 & ICM & 80.7190 & 33.3869 & 2.32 $\pm$ 0.12 & 2.30 $\pm$ 0.56 & -1.41 $\pm$ 0.56\\
 181173299780581120 & NGC 1893 & ICM & 80.7176 & 33.3843 & 2.35 $\pm$ 0.15 & 2.32 $\pm$ 0.35 & -1.73 $\pm$ 0.35\\
181173334136868224 & NGC 1893 & ISM & 80.6886 & 33.3713 & 0.86 $\pm$ 0.01 & 1.63 $\pm$ 0.26 & -0.34 $\pm$ 0.26\\
181173265420842880 & NGC 1893 & ISM & 80.6992 & 33.3686 & 0.94 $\pm$ 0.02 & 1.85 $\pm$ 0.43 & -0.33 $\pm$ 0.43\\
180986009142382720 & NGC 1893 & ICM & 80.7413 & 33.3687 & 2.68 $\pm$ 0.78 & 2.31 $\pm$ 0.56 & -2.18 $\pm$ 0.56\\
180985768624207744 & NGC 1893 & ICM & 80.7717 & 33.3593 & 2.87 $\pm$ 0.16 & 1.69 $\pm$ 0.72 & -1.90 $\pm$ 0.72\\
181178389317231616 & NGC 1893 & ICM & 80.6221 & 33.5140 & 3.01 $\pm$ 0.21 & 1.29 $\pm$ 0.14 & -2.99 $\pm$ 0.14\\
181175017764036352 & NGC 1893 & ISM & 80.6993 & 33.4761 & 0.84 $\pm$ 0.01 & 1.85 $\pm$ 0.24 & -1.07 $\pm$ 0.24\\
\hline
\end{tabular}

\vspace{0.5cm}

\begin{tabular}{ c c c c c c c c c }
\hline
$\mu_{RA}$ $\pm$ $\sigma$ & $\mu_{Dec}$ $\pm$ $\sigma$ & $P/d$ $\pm$ $\sigma$ & $A_{\mathrm{V}}$ $\pm$ $\sigma$  &  $P/A_{\mathrm{V}}$ $\pm$ $\sigma$   & $A_{\mathrm{Vint}}$ $\pm$ $\sigma$ & $P_{int}/A_{\mathrm{Vint}}$ $\pm$ $\sigma$  \\
\hline
(mas/yr) & (mas/yr)  & (\%/kpc) & (mag) &  (\%/mag) & (mag) & (\%/mag)   \\
    \hline
 -0.43 $\pm$ 0.02 & -1.53 $\pm$ 0.02 & 0.8 $\pm$  0.1 &1.67 $\pm$ 0.12 & 1.49 $\pm$ 0.16 & 1.35 $\pm$ 0.32 & 0.69 $\pm$ 1.92  \\
 -0.38 $\pm$ 0.02 & -1.48 $\pm$ 0.02 & 1.1 $\pm$ 0.3 &0.79 $\pm$ 0.02 & 3.19 $\pm$ 0.72 & 0.47 $\pm$ 0.29 & 1.04 $\pm$ 4.02 \\
 -0.29 $\pm$ 0.02 & -1.42 $\pm$ 0.02 & 1.2 $\pm$ 0.2 & 0.79 $\pm$ 0.02 &3.50$\pm$ 0.73 & 0.47 $\pm$ 0.29 & 1.28 $\pm$ 4.03\\
 -0.28 $\pm$ 0.03 & -1.27 $\pm$ 0.03 & 1.2 $\pm$ 0.2 & 0.79 $\pm$ 0.02 & 3.75 $\pm$ 0.47 & 0.47 $\pm$ 0.29 & 1.65 $\pm$ 4.86 \\
 1.29 $\pm$ 0.02 & -3.93 $\pm$ 0.03 & 1.9 $\pm$ 0.3 &0.63  $\pm$ 0.10& 2.72 $\pm$ 0.60 &  - & - \\
 -0.97 $\pm$ 0.02 & -8.04 $\pm$ 0.02 & 2.0  $\pm$ 0.5 & 0.63 $\pm$ 0.10 & 3.07 $\pm$ 0.85 &  - & -\\
 -1.00 $\pm$ 0.24 & -0.80 $\pm$ 0.25 & 1.2 $\pm$ 0.4 & 1.07 $\pm$ 0.09 & 3.06 $\pm$ 0.60 & 0.74 $\pm$ 0.31 & 1.50 $\pm$ 3.54 \\
 -0.52 $\pm$ 0.02 & -2.10 $\pm$ 0.02 & 0.9 $\pm$ 0.3 & 1.24 $\pm$ 0.11& 2.11 $\pm$ 0.63 & 0.92 $\pm$ 0.31 & 0.73 $\pm$ 3.14\\
 -0.90 $\pm$ 0.02 & -0.68 $\pm$ 0.02 & 1.1 $\pm$ 0.1 & 1.07 $\pm$ 0.09 & 3.14 $\pm$ 0.31 & 0.74  $\pm$ 0.31 & 2.44 $\pm$ 3.81 \\
 -3.15 $\pm$ 0.02 & -0.81 $\pm$ 0.02 & 2.6 $\pm$ 0.3 & 0.68 $\pm$ 0.02 & 3.23 $\pm$ 0.37 & - & -  \\
\hline
\end{tabular}

\label{tab:NGC1893}
\end{table*}
\subsection{\textbf{$\mathbf{P}/\mathbf{A}_{\mathrm{\mathbf{V}}}$ \textbf{\textit{versus}} $\mathbf{A}_{\mathrm{\mathbf{V}}}$ of ISM tracers}\label{3.3}}

\begin{figure}
\captionsetup[subfloat]{labelformat=empty}
\subfloat[\label{fig:pbyavism}]{\includegraphics[width=3.7in,height=3.125in]{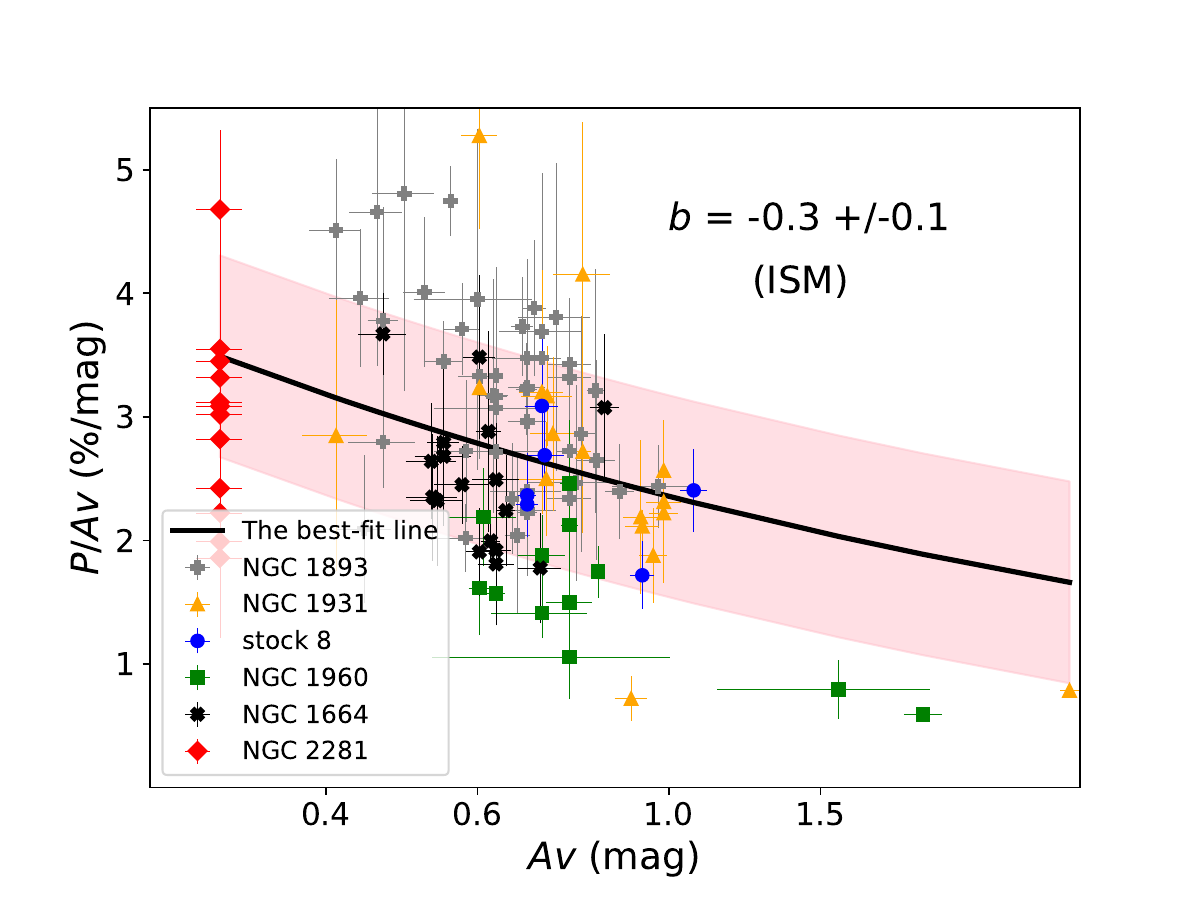}}\hspace{1em}\\
\caption{$P/A_{\mathrm{V}}$ {\it versus} $A_{\mathrm{V}}$   plot in the ISM. The best-fit power law is denoted with a thick line, and the corresponding 1-$\sigma$ confidence regions are shown as the shaded area in the background. The best-fit power-law index is also overlaid.}
\label{ismplots}
\end{figure}

The relation $P/A_{\mathrm{V}}$~--~$A_{\mathrm{V}}$ of the ISM dust towards the six clusters are fitted with a single power-law of the form $y=  a\left(x\right)^{b}$ using $scipy-curvefit$ module in {\sc Python}, where x is $A_{\mathrm{V}}$, $a$ is a coefficient, and $b$ is the power-law index. The best-fit parameters and the plot for $P/A_{\mathrm{V}}$~--~$A_{\mathrm{V}}$   relation towards ISM of the six stellar clusters are given in Table \ref{tab:brokenpowerlawparams} and Figure \ref{ismplots}  respectively. Based on the single power-law fit over $P/A_{\mathrm{V}}$~--~$A_{\mathrm{V}}$  data of ISM,  we infer that the $P/A_{\mathrm{V}}$ declines with $A_{\mathrm{V}}$ with an index of $-$0.3$\pm$ 0.1.  A sample data table containing the extinction ($A_{\mathrm{V}}$) and polarization efficiencies ($P/A_{\mathrm{V}}$) of the ISM tracers is given in Table \ref{tab:NGC1893}.

\subsection{Intrinsic \textbf{$\mathbf{P}$ and $\mathbf{A}_{\mathrm{\mathbf{V}}}$ of ICM tracers}}\label{3.4}



The polarisation and extinction values of ICM tracers (cluster members and background stars) can not be used directly to infer the properties of ICM dust. This is because they still contain the contribution from foreground dust; therefore, it is essential to quantify and subtract the same to shed light on the ICM dust. 
To do this, we extract the Stokes parameters $Q$ and $U$ of ISM tracers using the relations $Q = P \cos(2PA)$ and $U = P \sin(2PA)$, where $P$ and $PA$ are the fraction of polarisation and position angle of B-field inferred by ISM tracers.

\begin{figure}
\captionsetup[subfloat]{labelformat=empty}
\subfloat[\label{fig:Qismvsdistance}]{\includegraphics[width=3.7in,height=4in]{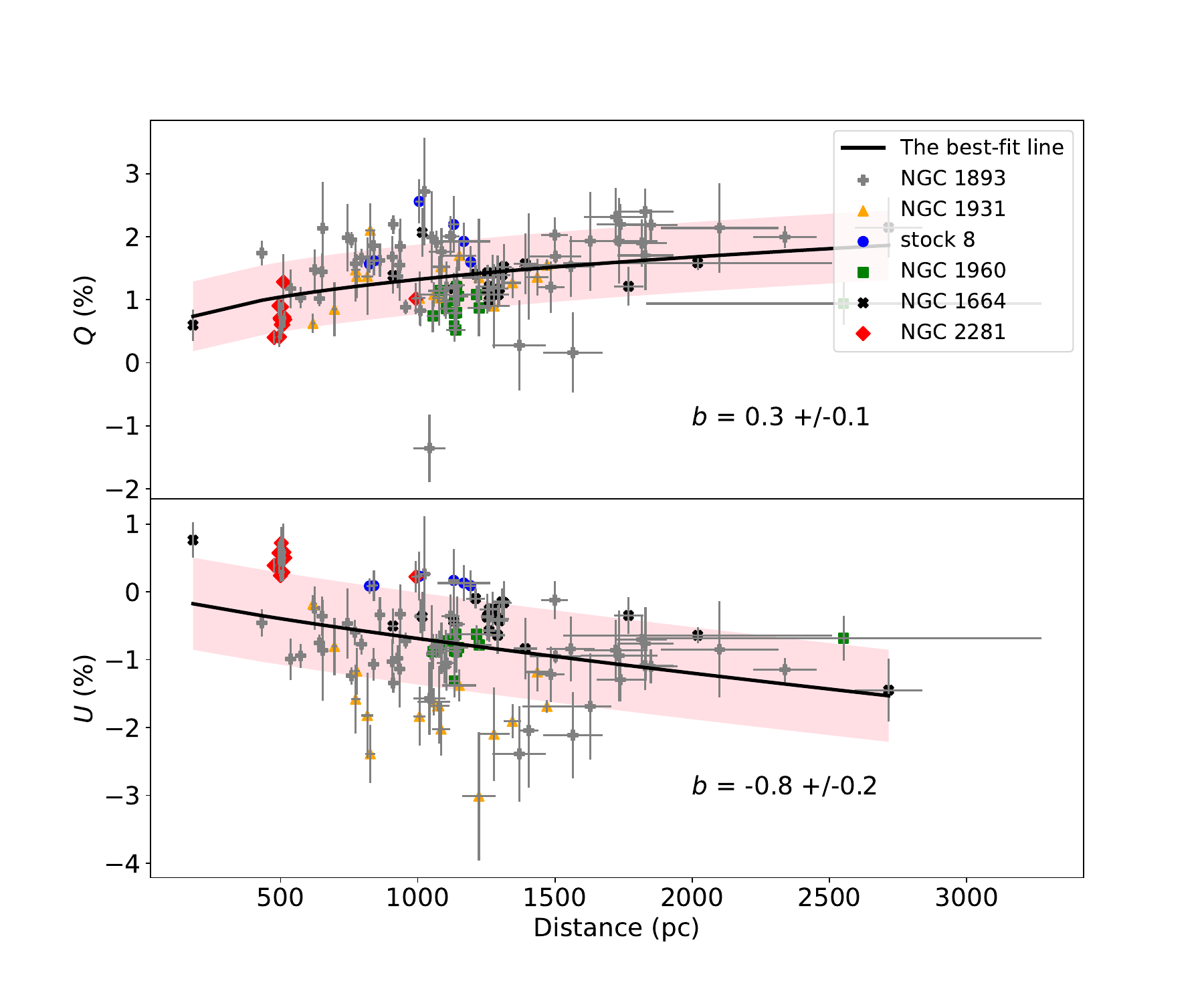}}\hspace{1em}\\
\caption{  $Q$  ({\it top}) and $U$ ({\it bottom}) {\it versus} distance plots of the ISM tracers (foreground stars of both young and old clusters as well as cluster members of old clusters). In both plots, Stokes parameters of six stars belonging to NGC\,1664 have been removed because they fell off from the main distribution and did not affect the overall fit.  The power-law fit is denoted with a thick line. The Best-fit parameters are also overlaid. The corresponding 1-$\sigma$ confidence regions are shown as shaded regions in both panels.} 
\label{stokesplots}
\end{figure}

Further, we perform power-law fits (similar to as described in section \ref{3.3}) on the Stokes parameters versus distance plots shown in Figure \ref{stokesplots} by considering the data of ISM tracers.  From the solution,  we then estimate the resultant Stokes parameters at the transition distance for each young cluster identified from GMM or DBSCAN, which are treated as foreground Stokes parameters, $Q_{\mathrm{\mathrm{fg}}}$ and $U_{\mathrm{fg}}$, and are listed in Table \ref{tab:foregroundparameters}. These are vectorially subtracted from the Stokes parameters of ICM tracers to obtain the intrinsic Stokes parameters for each young cluster using the following relations 
\begin{eqnarray}\label{qintuint}
                     Q_{\mathrm{int}} = Q - Q_{\mathrm{fg}}\nonumber\\ 
                     \mathrm{and}\,\,
                     U_{\mathrm{int}} = U - U_{\mathrm{fg}}.
\end{eqnarray}
Finally, the intrinsic fraction of polarisation corresponding to ICM dust is derived using 
\begin{eqnarray}\label{eq14}
 P_{\mathrm{int}}&=& \sqrt{ Q_{\mathrm{int}}^2+ U_{\mathrm{int}}^2}.\nonumber\\
\end{eqnarray}

\begin{figure}
\captionsetup[subfloat]{labelformat=empty}
\subfloat[\label{fig:avvsreddeningstock8}]{\includegraphics[width=3.45in,height=2.9in]{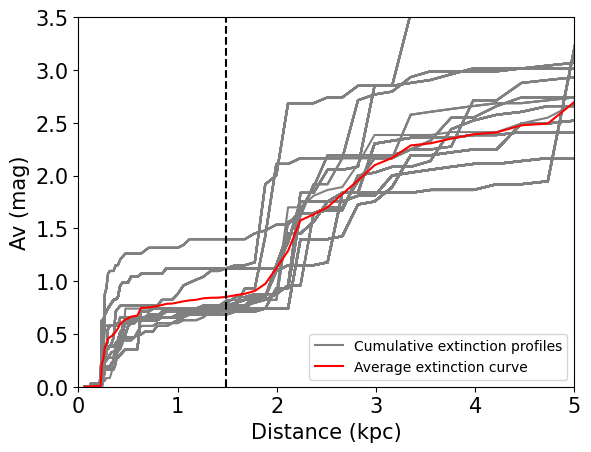}}\hspace{1em}\\
\caption{ Cumulative $A_{\mathrm{V}}$  {\it versus} distance profiles along all lines of sights in the 4.6\arcmin radius area around Stock\,8. The average extinction profile is also shown. The corresponding transition distance identified by the GMM is denoted by the vertical dotted line.} 
\label{reddeningcurvesstock8}
\end{figure}

Similarly, to determine foreground extinction towards three clusters Stock\,8, NGC\,1931, and NGC\,1893, we extract distance versus cumulative extinction, $A_{\mathrm{V}}$, profiles within a circular radius of  4.6\arcmin, 7.1\arcmin, and 10.5$\arcmin$ area, respectively, around each cluster. For this, we make use of the 3D extinction map of \citet{green2019} and {\sc dustmaps} module in {\sc python}. Figure \ref{reddeningcurvesstock8} depicts the distance versus extinction profiles for Stock 8. We take the average extinction from multiple extinction values at each distance and plot the average extinction profile (red curve) in Fig. \ref{reddeningcurvesstock8}. The mean $A_{\mathrm{V}}$ at the transition distance is considered the foreground extinction $A_{\mathrm{Vfg}}$ towards each cluster and is listed in Table \ref{tab:foregroundparameters}. 
We then subtract this $A_{\mathrm{Vfg}}$ from the $A_{\mathrm{V}}$ values of the ICM tracers by utilizing the relation $A_{\mathrm{Vint}} = A_{\mathrm{V}}- A_{\mathrm{Vfg}}$.



\begin{table*}
\caption{Best fit parameters, based on the broken power-law of overall $P/A_{\mathrm{V}}$~--~$A_{\mathrm{V}}$  and $P/d$~--~$d$ towards all the clusters,  single power-law of  $P/A_{\mathrm{V}}$~--~$A_{\mathrm{V}}$ towards the ISM, and single power-law of  $P_{\mathrm{int}}/A_{\mathrm{Vint}}$~--~$A_{\mathrm{Vint}}$ towards the ICM, respectively.}
\label{tab:brokenpowerlawparams}
\begin{tabular}{lccccccccl}

 \hline
 
Cluster-ID & Relation&  a & b & $b_{1}$ & $b_{2}$ &  $A_{\mathrm{V}}^{b}$ or $d^{b}$ \\
 \hline

Overall relations &  $P/A_{\mathrm{V}}$~--~$A_{\mathrm{V}}$ & 3.0 $\pm$ 0.1 & - & -0.03 $\pm$ 0.08  & -0.8 $\pm$ 0.1 &  0.9 $\pm$ 0.1 mag &\\

    &$P/d$~--~$d$ & 0.0015 $\pm$ 0.0007 & - & -0.5 $\pm$ 0.1 & -0.4 $\pm$ 0.1 &  1.3 $\pm$ 1.2 kpc\\
 \hline
ISM & $P/A_{\mathrm{V}}$~--~$A_{\mathrm{V}}$ & 2.3 $\pm$ 0.1 &-0.3 $\pm$ 0.1 & - & - & -  \\
 \hline
ICM & $P_{\mathrm{int}}/A_{\mathrm{Vint}}$~--~$A_{\mathrm{Vint}}$ & 1.4 $\pm$ 0.1 &-0.7 $\pm$ 0.1 & - & - & -  \\
 \hline
\end{tabular}
\end{table*}

\begin{table*}
\caption{The foreground polarisation and extinction values of the three young clusters in the anti-center Galaxy direction}
\label{tab:foregroundparameters}
\begin{tabular}{lccccccl}
 \hline
 Cluster-ID  & $Q_{\mathrm{\mathrm{fg}}}$ & $U_{\mathrm{\mathrm{fg}}}$ &  $A_{\mathrm{Vfg}}$ \\
\hline
 & (\%) & (\%) & (mag) \\
\hline
Stock\,8   &1.66&$-$1.12 & 0.41 \\
NGC\,1931  & 1.62 &$-$1.05& 0.34 \\
NGC\,1893 & 1.73&$-$1.23 &0.33 \\
\hline
\end{tabular}
\end{table*}

\subsection{$\mathbf{P_{\mathrm{\mathbf{int}}}}/\mathbf{A_{\mathrm{\mathbf{Vint}}}}$ \textbf{\textit{versus}} $\mathbf{A_{\mathrm{\mathbf{Vint}}}}$ relations towards the ICM}\label{3.5}

To examine the trends in the dust polarisation efficiency in the ICM using the foreground subtracted $P_{\mathrm{int}}$ and $A_{\mathrm{Vint}}$ (cf., Section \ref{3.4}),  the $P_{\mathrm{int}}/A_{\mathrm{Vint}}$~--~$A_{\mathrm{Vint}}$ relations towards the three younger clusters Stock\,8, NGC\,1931, and NGC\,1893 are plotted and are fitted with a power-law. The $P_{\mathrm{int}}/A_{\mathrm{Vint}}$~--~$A_{\mathrm{Vint}}$ plots with the corresponding fit for the ICM are shown in figure \ref{icmoldplots}. From the figure, we see that the $P_{\mathrm{int}}$/$A_{\mathrm{Vint}}$ in the ICM decreases with a power-law index of $-$0.7 $\pm$ 0.1 . The best-fit parameters are given in Table \ref{tab:brokenpowerlawparams}.  A sample data table containing the intrinsic extinction ($A_{\mathrm{Vint}}$) and intrinsic polarisation efficiencies ($P_{\mathrm{int}}/A_{\mathrm{Vint}}$) of the ICM tracers is given in Table \ref{tab:NGC1893}.

\section{Discussion}\label{section4}

The power-law fits on $P/A_{\mathrm{V}}$~--~$A_{\mathrm{V}}$ and $P_{\mathrm{int}}/A_{\mathrm{Vint}}$~--~$A_{\mathrm{Vint}}$  of ISM and ICM tracers suggest different power-law indices of $-0.3$ and $-0.7$, respectively. 
These indices imply efficient dust grain alignment in the general diffuse ISM than those dust grains in the ICM.  This is further evidenced by the fact that, despite having a smaller mean $A_{\mathrm{V}}$ of 0.7 $\pm$ 0.3 mag, the ISM exhibits relatively a higher mean $P/A_{\mathrm{V}}$ of 2.7 $\pm$ 0.9 \% mag$^{-1}$. In contrast, the ICM with a higher mean $A_{\mathrm{Vint}}$ of 1.1 $\pm$ 0.6 mag shows a smaller $P_{\mathrm{int}}/A_{\mathrm{Vint}}$ of 1.5 $\pm$ 1.0 \% mag$^{-1}$. Even though different mechanisms have been proposed to explain how the dust grains are getting aligned with the magnetic fields to produce higher polarisation, the most established has been the Radiative Torque (RAT) alignment mechanism \citep{dolginov1976,draine1996,LazarianHoang2007}.  According to the RAT mechanism, when a radiation field is incident upon the dust grains, it spins the dust grains to higher angular momentum and hence can efficiently align them with the local magnetic fields. These efficiently aligned dust grains then cause a higher level of polarisation. Therefore, higher radiation field strength can result in better dust grain alignment and increased polarisation. Similarly, the size of the dust grains is another important factor determining the extent of dust alignment. If the dust grains are too large, RATs will become inefficient in rotating them. On the other hand, if the dust grains are too small, they may be spherical to scatter radiation anisotropically and generate RATs. As a result, only larger sub-micron-sized dust grains that are neither too big nor too small tend to align well with magnetic fields \citep{whittet2004}.

The diffuse ISM receives a higher amount of interstellar radiation field, which can result in stronger RATs that rotate the dust grains to higher angular momentum, leading to better alignment with the local magnetic field and producing a higher fraction of interstellar polarisation as seen in Fig. \ref{ismplots}. To explain the poor alignment of dust grains observed in the ICM compared to ISM, it is essential to probe the intracluster medium conditions of the three young clusters: Stock\,8,  NGC\,1931, and NGC\,1893. 

\begin{figure}
\captionsetup[subfloat]{labelformat=empty}
\subfloat[\label{fig:pbyavyoungclust}]{\includegraphics[width=3.8in,height=3.3in]{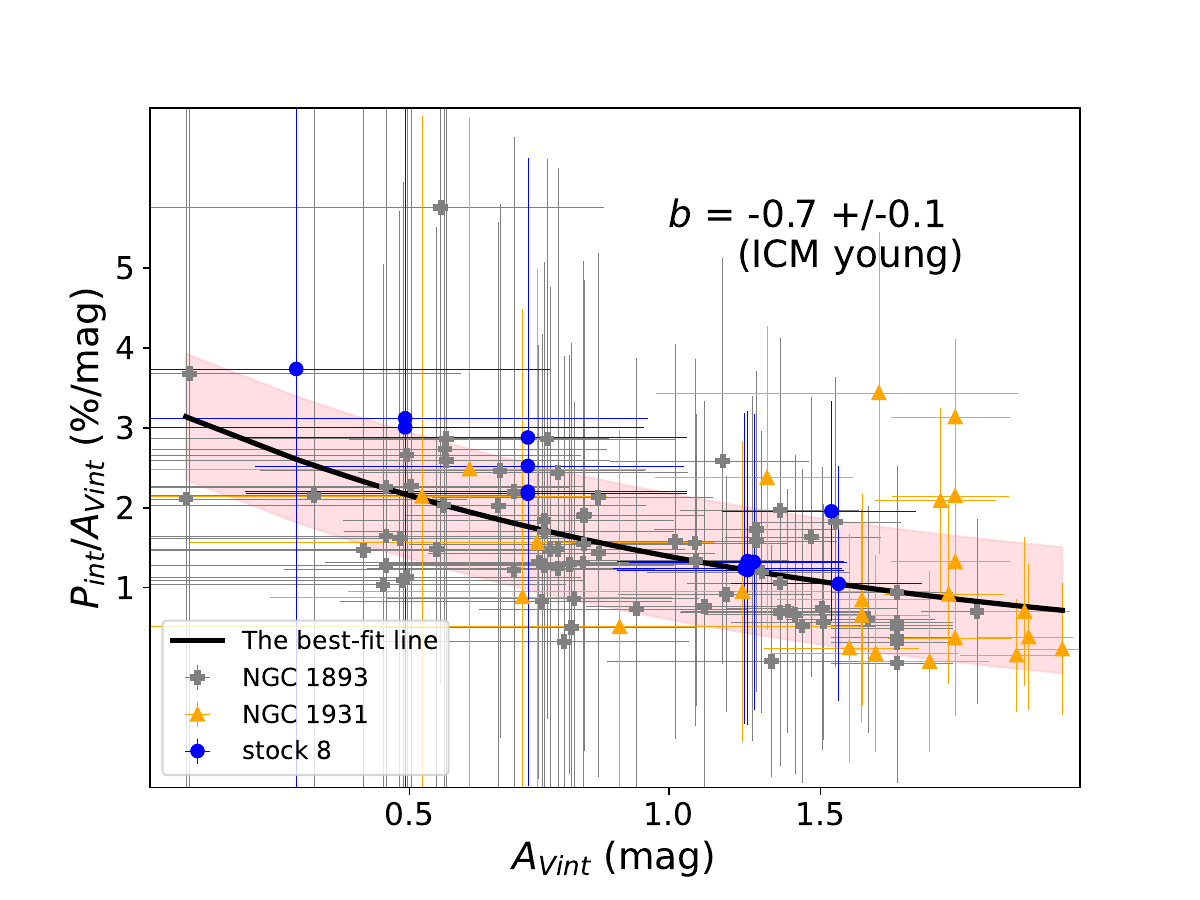}}
\caption{$P_{\mathrm{int}}/A_{\mathrm{Vint}}$ {\it versus} $A_{\mathrm{Vint}}$  plot in the ICM. The best-fit power-law is denoted with the thick line, and the corresponding 1-$\sigma$ confidence regions are shown as the shaded area in the background. The best-fit power-law index is also overlaid.} 
\label{icmoldplots}
\end{figure}

Out of the three clusters, NGC\,1893 is the youngest followed by Stock 8 and NGC\,1931 respectively, with NGC\,1893 having the highest number of O/B type stars followed by Stock\,8 and NGC\,1931 \citep{Bonatto2009,damian2021,alexquitana2023}. As shown in figure \ref{icmoldplots}, NGC\,1893 exhibits relatively higher polarisation efficiency ($P_{\mathrm{int}}/A_{\mathrm{Vint}}$) among the three clusters, followed by Stock\,8 and NGC\,1931. This difference in polarisation efficiencies can be attributed to the varying number of O/B type stars, as a higher number of such stars can produce more energetic UV radiation, leading to stronger RATs that efficiently align the dust grains and produce higher polarisation in NGC\,1893 as compared to the other two clusters.  Similarly, NGC\,1931 contains relatively bigger-sized dust grains as inferred by the abnormal reddening law observed in the ICM of NGC\,1931 with a $R_{\mathrm{V}}$ value of 5.2 $\pm$ 0.3 \citep{Beomdulim2015}. In addition, \citet{jessyjose2008} has found the presence of small-sized dust grains and small polycyclic aromatic dust grains in the intracluster medium of Stock\,8 using the MIR emission based on Midcourse Space Experiment (MSX).

The conditions in the ICM of younger clusters, as described above, can have several consequences for the alignment of dust grains with magnetic fields and the resulting polarisation efficiencies. For instance, all three young clusters, especially NGC\,1893  have strong radiation feedback and ionization shocks emanating from the massive O/B type stars which heat the gas in the ICM \citep{jessyjose2008,pandeyetal2013,bomdulim2018,damian2021}. This increased gas temperature in the ICM of the younger clusters may be causing increased gas-dust collisions, which might be misaligning the dust grains in the ICM.  Similarly, due to bigger and smaller-sized dust grains in the ICM of NGC\,1931 and Stock\,8, respectively, RATs can become ineffective in aligning the dust grains with the magnetic field. Hence, a combined effect of all these possible scenarios in action can explain the overall lower polarisation efficiency of dust grains seen in the ICM vis-\`a-vis ISM.  This demonstrates that our polarisation data can trace the POS magnetic fields and dust properties of ISM dust very well owing to efficient dust grain alignment, but does poorly in the ICM dust of young clusters due to relatively inefficient dust grain alignment.

\begin{figure}
\captionsetup[subfloat]{labelformat=empty}
\subfloat{\includegraphics[width=3.8in,height=4.125in]{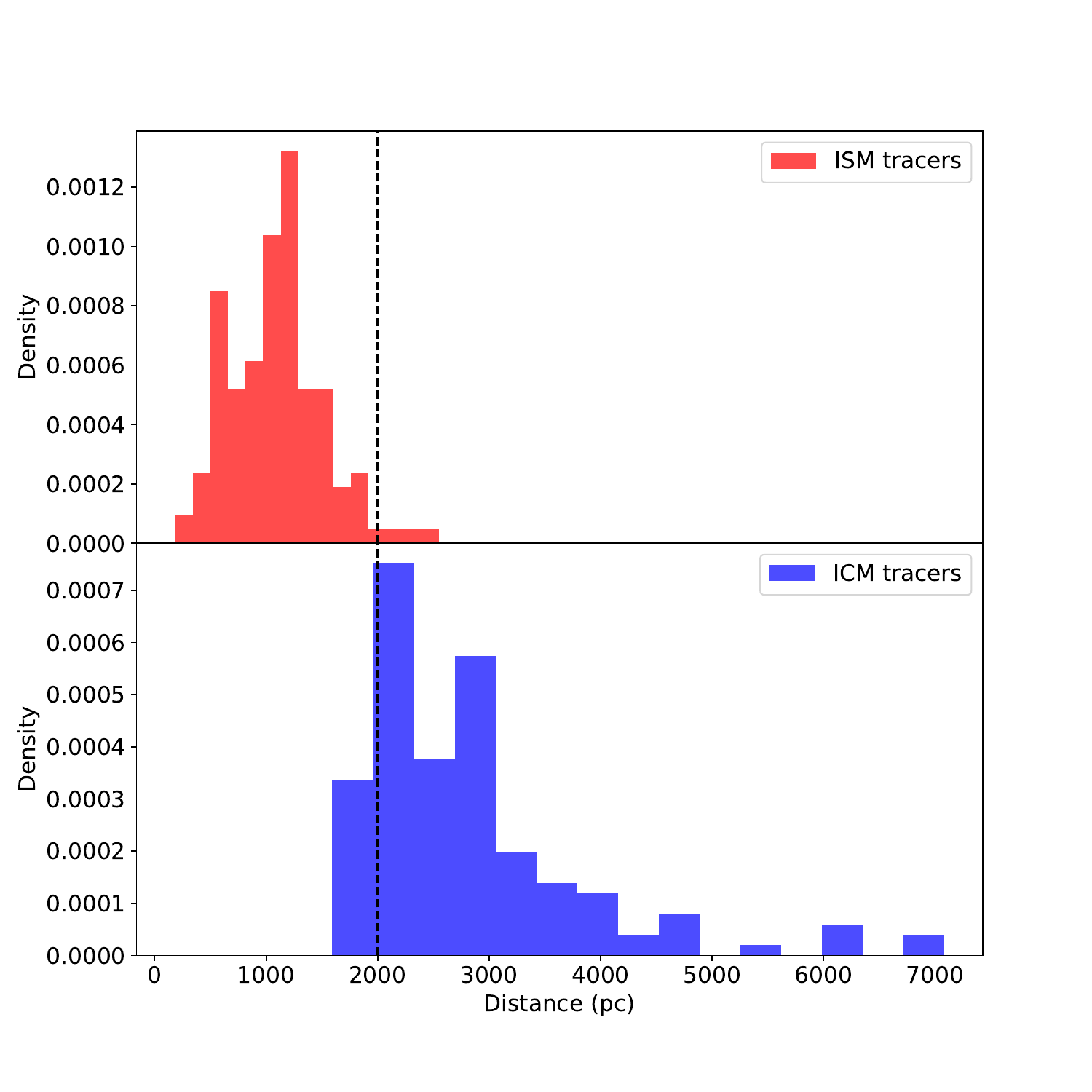}}
\caption{Histograms of the stars tracing the ISM dust ({\it top}) and ICM dust ({\it bottom}). The dotted vertical lines in both figures denote the transition distance of 2 kpc.} 
\label{ismandicmhistograms}
\end{figure}

As per the distribution of distances shown in Figure \ref{ismandicmhistograms}, it is evident that the ISM tracers are confined up to 2 kpc, whereas the ICM tracers lie beyond 2 kpc. We have witnessed this distance boundary of 2 kpc where ISM and Perseus dust have been apportioned towards the direction of NGC\,1893 \citep{Bijas2022}. Therefore, the current study based on the data of several clusters confirms that there exist two dust populations that are separated at 2 kpc. It is worth noting here that the proposed Perseus dust by  \citet{Bijas2022} and the termed ICM dust in this work should be the same because the stars lying beyond 2 kpc must either be part of young clusters or background stars tracing ICM dust.

\begin{figure}
\captionsetup[subfloat]{labelformat=empty}
\subfloat[\label{ismandicmhist}]{\includegraphics[width=3.5in,height=3in]{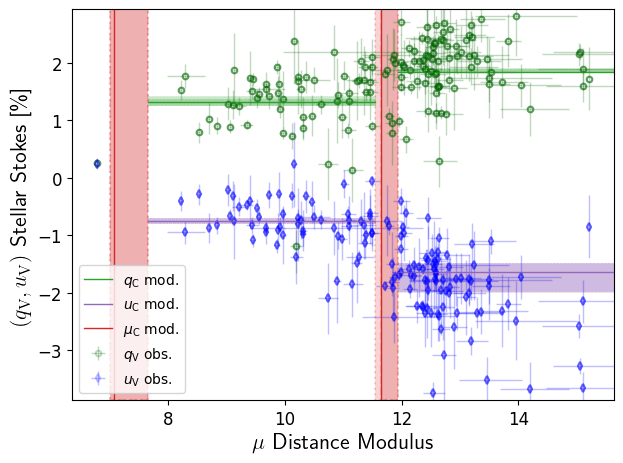}}
\caption{The stellar Stokes parameters $q_\mathrm{{V}}$ (circles) and $u_\mathrm{{V}}$ (diamonds) vs distance modulus ($\mu$) plot from BISP-1 two dust layer model. The horizontal lines denote the posterior mean Stokes parameters $q_\mathrm{{C}}$ mod. and $u_\mathrm{{C}}$ mod. of each cloud, respectively. The vertical lines denote the mean distance moduli corresponding to the mean posterior parallaxes of the two clouds. In each case,  the corresponding shaded regions indicate the 95 percentile (light shade) and 68 percentile (dark shade) confidence regions.} 
\label{bisp_1 plot}
\end{figure}
\begin{figure*}
\captionsetup[subfloat]{labelformat=empty}
\subfloat[\label{ismandicmhist}]{\includegraphics[width=5.75in,height=7in]{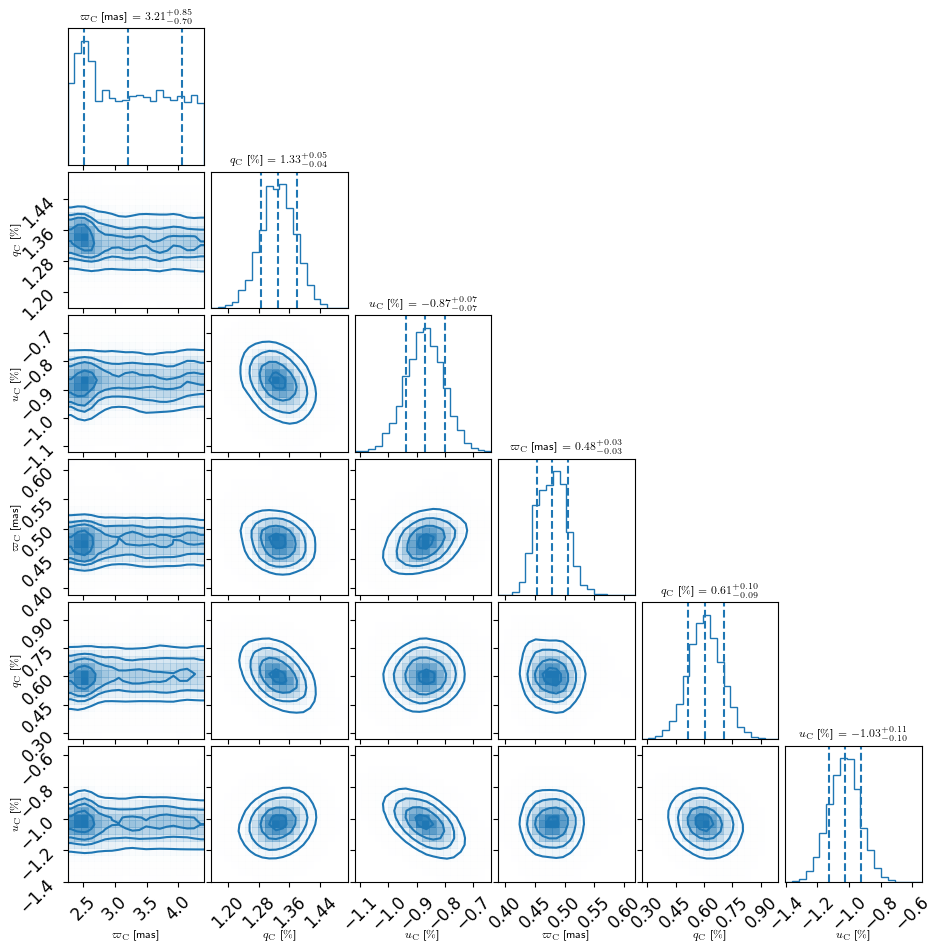}}
\caption{The posterior distributions of the cloud parallaxes ($\omega_{C}$, in units of milliarcseconds) and mean Stokes parameters ($q_\mathrm{{C}}$ and $u_\mathrm{{C}}$, in units of \%) for the 2 dust layer model produced by BISP-1. The vertical dashed lines indicate 16, 50, and 84 percentiles of each posterior distribution. The values at the top of each panel contain the mean, minimum, and maximum values of each parameter corresponding to the 68 \% confidence interval.}
\label{bisp1_posteriors}
\end{figure*}

To confirm the 2 kpc boundary between two dust components of ISM and ICM, and to quantify the relative polarisation contribution of each component independently,
we used the recently published {\sc python} package BISP-1\footnote{\url{https://github.com/vpelgrims/Bisp_1/}} \citep{pelgrims2023}. 
BISP-1 uses an LOS inversion method to decompose the observed polarisation into individual polarisation contributions from each dust cloud along the LOS. We gave the observed Stokes parameters ($q_\mathrm{{V}}$, $u_\mathrm{{V}}$), parallaxes, and their corresponding errors of all the stars towards the six clusters as input to the BISP-1 package. A two-dust layer model was invoked to predict the posterior mean Stokes parameters ($q_\mathrm{{C}}$ mod. and $u_\mathrm{{C}}$ mod.) and distance moduli ($\mu_\mathrm{{C}}$ mod.) of the individual dust clouds contributing to the overall polarisation. The two-dust layer model consisted of 12 flat prior distributions (2 for each parameter corresponding to the minimum and maximum values), and a dynesty sampler was used to sample the parameter space to produce the posterior distributions using 50000 sampling points. Details on how the two-dust layer model was chosen as the best model are given in Appendix \ref{bisp_modelsel}. The resulting Stokes parameters versus distance modulus plot are shown in Figure \ref{bisp_1 plot}, and the corresponding posterior distributions of each cloud's parallaxes and Stokes parameters are given in Figure \ref{bisp1_posteriors}.  From both figures, we see that BISP-1 predicts two dust layers: (i) the first dust layer tracing the foreground medium starts at a distance of $\sim$0.31 kpc ($\mu$ = 7.44) and (ii) ICM dust appearing from 2.1 kpc onwards ($\mu$ = 11.5) traces the intracluster medium. Using the mean posterior values of the Stokes parameters of the individual clouds ($q_\mathrm{{C}}$ and $u_\mathrm{{C}}$), we estimated the percentage polarisation for each. It was seen that the ISM dust cloud has a higher percentage of polarisation of 2.3\% and is a major contributor to the overall polarisation than the ICM dust cloud, with a polarisation of 1.5\%. 
Note that the BISP-1-yielded ICM contribution takes care of the foreground contribution. Therefore, ISM dust contributes a higher amount of polarisation than that of ICM. 


We also estimated the mean polarisation angles of the two dust clouds using the Stokes parameters, which were found to be 163.4$\degr$ and 150.1$\degr$,  respectively. The similar polarisation angles of the ISM and ICM dust clouds indicate that the B-field in the ISM and ICM is nearly the same. 
This means that the changes observed in the polarisation efficiencies in the ISM and ICM are not due to a change in the orientation of B-fields along the LOS of target clusters, but due to differences in the properties of ISM and ICM dust. This result, therefore, also supports results of \citet{Bijas2022}, which have also reported higher alignment efficiency of foreground dust up to 2 kpc and poor alignment efficiency of Perseus dust after 2 kpc despite having uniform B-field orientation towards NGC\,1893.

\section{Summary and Conclusions}\label{section5}
Our study aims to understand the changes in dust polarisation efficiencies in the ISM and ICM using the polarisation data of three older and three younger clusters. These clusters are spread over a 10-degree large area towards the anti-center galaxy. To achieve this, we categorize the stars of each cluster into those tracing ISM and ICM dust using two clustering algorithms GMM and DBSCAN. We then analyzed the variations in the polarisation efficiency as a function of extinction in (i) diffuse ISM, and (ii) ICM of younger clusters by utilizing $V$-band polarisation data from the ARIES IMaging POLarimeter (AIMPOL) \citep{eswaraiah2011,pandeyetal2013,Bijas2022}, distances from \citet{bailerjones2021}, and extinction data from \citet{green2019}. Our findings suggest more efficiently aligned dust grains in the ISM than those in the ICM of younger clusters, possibly due to the unfavourable dust sizes and physical conditions in the ICM. We also found that ISM and ICM dust are separated at 2 kpc distance which is in accordance with our previous findings \citep{Bijas2022}. Further, we have used the LOS inversion technique, BISP-1 \citep{pelgrims2023} to confirm the distance boundary of 2 kpc and to quantify the relative polarisation contribution of ISM and ICM dust. BISP-1 shows that ISM dust contributes a higher amount of fraction of polarisation in comparison to ICM dust. 
BISP-1 also reveals that the mean B-field orientation in the ISM and ICM is nearly identical.  

\section*{Acknowledgements}
We thank the referee, Dr. Ralf Siebenmorgen, for the insightful and constructive suggestions on the manuscript, which have improved the content and flow of this paper. C.E. acknowledges the financial support from grant RJF/2020/000071 as a part of the Ramanujan Fellowship awarded by the Science and Engineering Research Board (SERB), Department of Science and Technology (DST), Govt. of India. We thank Chi Thiem Hoang, Vincent Pelgrims, and Gina Panopoulou for the fruitful discussions. We also would like to thank Trupti Nayak for her suggestions in creating the schematic diagram of the six clusters.

\section*{Data Availability}
Partial data is given in Table \ref{tab:NGC1893}. The complete dataset can be accessed online from the Harvard Dataverse using the following link: \url{https://dataverse.harvard.edu/dataset.xhtml?persistentId=doi%3A10.7910%2FDVN%2FHHYYGK&version=DRAFT}



\bibliographystyle{mnras}
\bibliography{example} 




\appendix

\section{Model selection of two dust layer model using BISP-1}\label{bisp_modelsel}
To determine the number of dust layers along the LOS of six clusters, we used the BISP-1 package to fit the stellar polarisation and parallax data with one-dust layer and two-dust layer models. The best model was chosen based on two criteria (i) model evidence returned by the nested sampling method from the dynesty sampler and (ii) akaike information criterion (AIC), which measures the information loss when the model is used to represent the data. The best model is the one with the highest model evidence and the lowest AIC value. From the model comparisons, BISP-1 predicted the two-layer model as a better model than that of the one-dust layer. 
We tried fitting the data with a three-layer model too, but the results were similar. So, we decided to go with the two-layer model.


\bsp	
\label{lastpage}
\end{document}